\documentclass[a4paper,11pt]{article}
\usepackage{pos}
\usepackage{amsmath}
\usepackage{bm}
\graphicspath{{figs/}}

\newcommand{\Vcb}{|V_{cb}|}
\newcommand{\Vub}{|V_{ub}|}

\newcommand{\matrixel}[3]{\left< #1 \vphantom{#2#3} \right|
 #2 \left| #3 \vphantom{#1#2} \right>} % for Dirac matrix elements
\newcommand{\bra}[1]{\langle #1 |}
\newcommand{\ket}[1]{| #1 \rangle}

\title{$B$-meson semileptonic decays from highly improved staggered quarks}
\author*[a]{Andrew Lytle}
\author[b]{Carleton DeTar}
\author[a]{Aida El-Khadra}
\author[c]{Elvira G{\'a}miz}
\author[d]{Steven Gottlieb}
\author[e]{William Jay}
\author[f]{Andreas Kronfeld}
\author[g]{Jack Laiho}
\author[f]{James Simone}
\author[h]{Alejandro Vaquero}

\affiliation[a]{Department of Physics and Illinois Center for Advanced Studies of the Universe, University of Illinois, Urbana, Illinois, 61801, USA}
\affiliation[b]{Department of Physics and Astronomy, University of Utah, Salt Lake City, Utah 84112, USA}
\affiliation[c]{Departamento de Física Teórica y del Cosmos, Universidad de Granada, E-18071, Spain}
\affiliation[d]{Department of Physics, Indiana University, Bloomington, Indiana 47405, USA}
\affiliation[e]{Center for Theoretical Physics, Massachusetts Institute of Technology, Cambridge, MA 02139, USA}
\affiliation[f]{Theory Division, Fermi National Accelerator Laboratory, Batavia, Illinois, 60510, USA}
\affiliation[g]{Department of Physics, Syracuse University, Syracuse, NY  13244, USA}
\affiliation[h]{CAPA \& Departamento de Física Teórica, Universidad de Zaragoza, 50009 Zaragoza, Spain}

\emailAdd{atlytle@illinois.edu}

\abstract{We present an update for results on $B$-meson semileptonic decays using the highly improved staggered quark (HISQ) action for both valence and 2+1+1 sea quarks. The use of the highly improved
action, combined with the MILC collaboration’s gauge ensembles with lattice spacings down to $\sim$0.03
fm, allows the $b$ quark to be treated with the same discretization as the lighter quarks. The talk will
focus on updated results for $B_{(s)} \to D_{(s)}$, $B_{(s)} \to K$ scalar and vector form factors. \\

}

\FullConference{The 40th International Symposium on Lattice Field Theory (Lattice 2023)\\
July 31st - August 4th, 2023\\
Fermi National Accelerator Laboratory\\}

\begin{document}
\maketitle

%----------------------------------------------------------------------
\section{Introduction}
Lattice QCD calculations of meson-decay form factors are crucial inputs for high-precision tests of the Standard Model (SM) of particle physics.
Combined with the corresponding experimentally measured decay rates, the lattice 
determinations of hadronic
flavor-changing matrix elements allow for independent determinations of CKM matrix elements, enabling precision tests of the SM. Lattice form-factor data also give shape information, and pure SM predictions for quantities such as $R$-ratios.

Thanks to ongoing experimental programs for heavy flavor physics including Belle II, BES III, and those at the LHC, we expect new measurements for many different observables and decay channels with increasing precision in the coming years. Improvements on the SM theory side are therefore well motivated and timely, and
 will enable sharper
tests of the SM, provided theory uncertainties are quantified
at a commensurate level as experiment.
 Some current few-sigma discrepancies (collectively referred to as $B$-anomalies) may be hints for new physics, 
(For recent discussions from a lattice perspective, see, e.g., Refs.~\cite{Boyle:2022uba,USQCD:2022mmc,Davoudi:2022bnl,Vaquero:2022mak,Tsang:2023nay}.)
There are also long-standing differences in inclusive and exclusive extractions of the CKM elements $\Vub$ and $\Vcb$, which should be resolved. Improved theoretical calculations of $B$-semileptonic decays will bear directly both on exclusive/inclusive discrepancies, and interpreting $B$-anomalies. 
Here we provide a status update of the FNAL-MILC collaboration's calculations of semileptonic $B_{(s)}$-meson decay form factors using the highly improved staggered quark (HISQ) action.

 %----------------------------------------------------------------------
\section{Calculation overview}
Some of the details of our calculation have been reported previously in Refs.~\cite{Lytle:2023xuq,FermilabLattice:2021bxu}. Here we review the basics and highlight updates from previous reports.
Our calculation uses ensembles generated by the MILC Collaboration using $N_f=2+1+1$ flavors of dynamical sea quarks with the HISQ action~\cite{MILC:2010pul,MILC:2012znn,Bazavov:2017lyh}.
We report preliminary results from ensembles with lattice spacings of $a = 0.09$, 0.06, 0.042, and 0.03 fm. At $a = 0.09$ and 0.06 fm we have generated correlator data
on ensembles with light sea-quarks at their physical values as well as at $m_l/m_s = 0.1$ and  0.2.
At $a= 0.042$ and 0.03 fm we have analyzed ensembles with $m_l \approx 0.2 m_s$ in the sea, and are currently generating 0.042 fm data with physical light quarks in the sea.
The strange and charm sea-quark masses are tuned to be close to their physical values, and the valence light- and strange-quark masses are taken to be equal to the corresponding sea-quark masses.
The heavy valence quarks range in mass from roughly $0.9 m_c$ to near the lattice cutoff, $a m_h  \approx 1$.
At the finest lattice spacings, this setup allows simulation close to, or in the case of $a = 0.03$ fm directly at, the physical mass of the bottom quark.

\begin{figure}[t]
    \centering
    \includegraphics[width=0.47\textwidth]{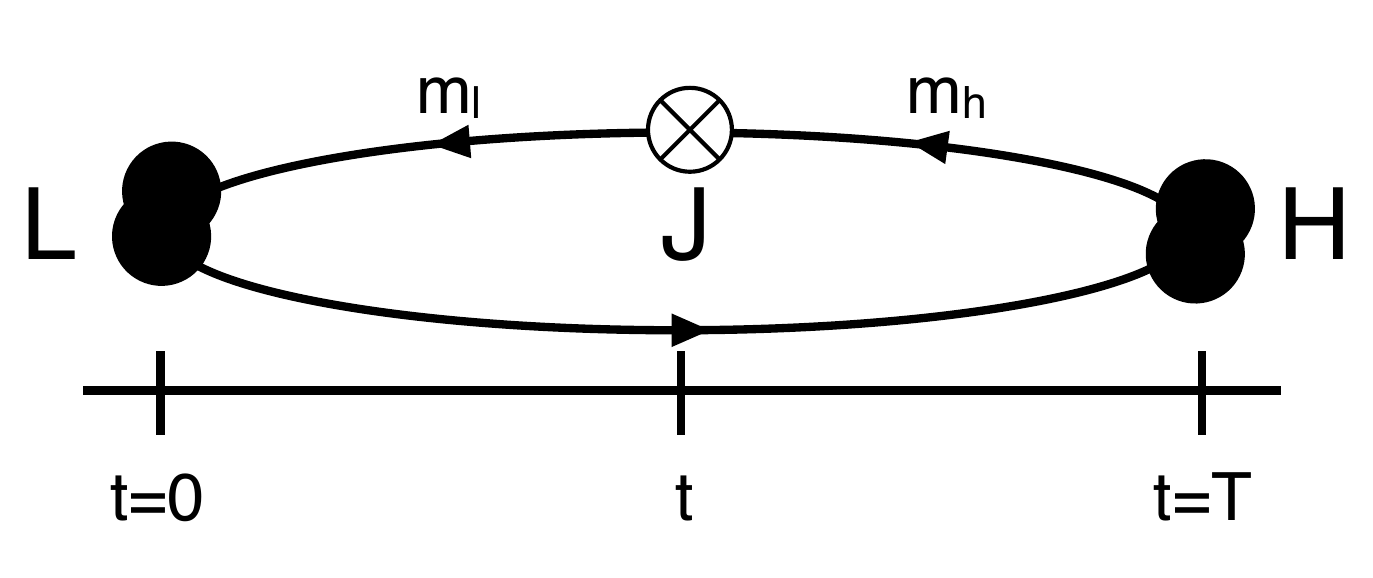}
    \caption{A schematic of the 3pt functions
    used to determine matrix elements entering Eqs.~\eqref{eq:f_parallel}--\eqref{eq:f_0}
    The light final state hadron is created with momentum $\bm{p}$ at the origin.
    An external current $J$ is inserted at time $t$.
    The heavy hadron is destroyed at rest at time $T$.
    }
    \label{fig:schematic_3pt}
\end{figure}

To determine the required form factors, we first extract matrix elements from joint correlated fits to two-point and three-point correlation functions.
More details of our correlators and fit forms are given in Ref.~\cite{Lytle:2023xuq}.
A schematic of a generic three point function used in our fits is shown in Fig.~\ref{fig:schematic_3pt}.
As usual for staggered fermions, the correlation functions include smoothly decaying contributions with the desired parity as well as oscillating contributions from states of opposite parity. We fit these correlators varying the number of even and odd states ($n$, $m$), checking stability in our final results to these variations.

All of our results are blinded by a random factor that is common for all three-point functions of a given analysis/decay channel.
We will carry the analysis of the blinded form factors all the way through the chiral interpolation and continuum extrapolation, unblinding only when the analysis of systematic errors is complete.

%----------------------------------------------------------------------
\section{Form factors}
Having extracted scalar and vector three-point matrix elements from simultaneous fits, we can relate these to  decay form factors via  
\begin{align}
f_\parallel	&= Z_{V^0}\frac{\matrixel{L}{V^0}{H}}{\sqrt{2 M_H}} \label{eq:f_parallel}\\
f_\perp		&= Z_{V^i}\frac{\matrixel{L}{V^i}{H}}{\sqrt{2 M_H}} \frac{1}{p^i_L} \label{eq:f_perp}\\
f_0				&= \frac{m_h-m_\ell}{M_H^2 - M_L^2} \matrixel{L}{S}{H}. \label{eq:f_0}
\end{align}
In these expressions, $M_H$, $M_L$, $p_H^\mu$, and $p_L^\mu$ refer to the mass and four-momentum of the heavy initial (H) and light final state (L) mesons; $m_h$ and $m_\ell$ refer to the heavy and light input quark masses of the transition current.
The final equality relating $f_0$ to the scalar matrix element follows from partial conservation of the vector current.

\begin{figure}
\centering
\includegraphics[width=0.58\textwidth]{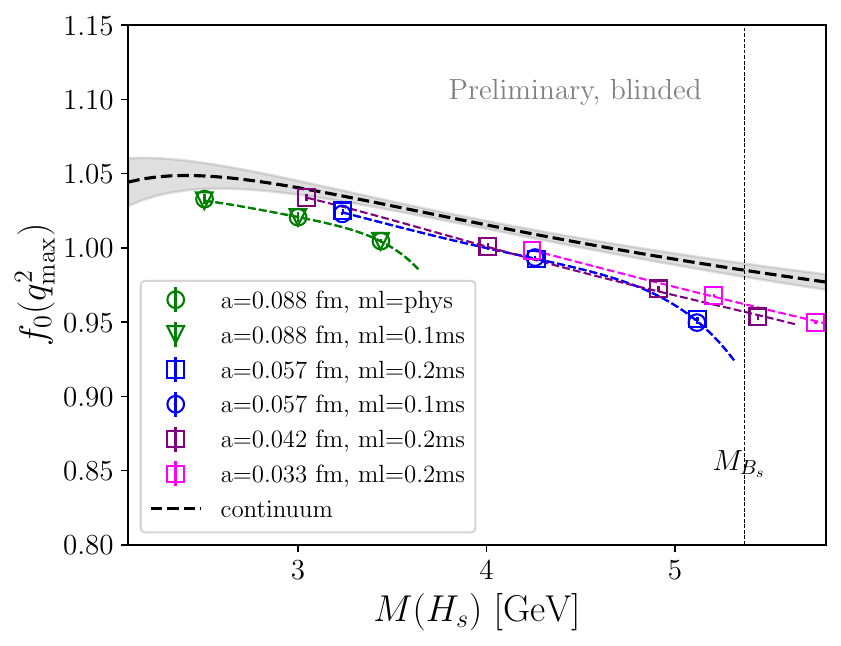}
\caption{
An example continuum fit for the $B_s \to D_s$ decay $f_0$ form factor, at the zero-recoil ($q^2_{\text{max}}$) point. Here the $x$ axis gives the mass of a proxy $H_s$ meson, the results at the physical $B_s$ mass are marked by the vertical dotted line. The symbols give the results from lattice data, while the dotted lines give the results of the fit evaluated for the given ensemble parameters. The black dashed line and gray band give the results of the fit evaluated in the $a \to 0$ limit. Note the data are blinded.
\label{fig:f0_q2max}}
\end{figure}

In Fig.~\ref{fig:f0_q2max}, we show results for the $H_s \to D_s$ scalar form factor $f_0$ at zero recoil, as a function of the heavy proxy meson mass $M(H_s)$. For this decay involving no light quarks and at zero-recoil, the data itself are very precise.
On our finest ensembles $a = 0.042$ fm and $a = 0.03$ fm, we have data points near to (and for $a=0.03$ fm beyond) the physical $B_s$ mass, indicated by a vertical dotted line in the figure. For the $a = 0.09$ fm and $a = 0.06$ fm data, we have results with two different light-quark masses in the sea. The extremely weak sea-quark mass dependence is evident here. We can perform a simple fit to the data, modeling the physical dependence as a polynomial in $1/M_{H_s}$ and discretization artifacts as a polynomial in $(am_h)^2$,
\begin{equation}
    f_0(q^{2}_\text{max})[M_{H_s}, am_h] = \sum_{ij} c_{ij} \Bigl(\frac{1}{M_{H_s}}\Bigr)^{i} \Bigl(a m_h\Bigr)^{2j} \,
\end{equation}
to determine the physical result for the $f_0(q^2_{\text{max}})$ form factor. The continuum result for the form factor as a function of $M(H_s)$ is given by the black dashed line.

At Lattice 2022, we showed raw data for the $B_s \to D_s$ decay $f_0$, $f_\parallel$, and $f_\perp$ form factors~\cite{Lytle:2023xuq}. Here, we focus on the analogous data relevant for the decay $B_s \to K$, obtained by using $h \to l$ currents instead of the $h \to c$ currents for the former decay. Raw data for these decays are given by the colored symbols in Fig.~\ref{fig:BstoK}. For this figure, we are displaying only data $m_h \lesssim 3.0 m_c$, and we discuss a preliminary chiral and continuum fit to the data in the next section. We observe good precision in the data, especially at lower recoil, though overall the data is noisier than for $B_s \to D_s$ on account of the light quark propagator.
%----------------------------------------------------------------------
\section{Vector current normalization}
\begin{figure}
\centering
\includegraphics[width=0.55\textwidth]{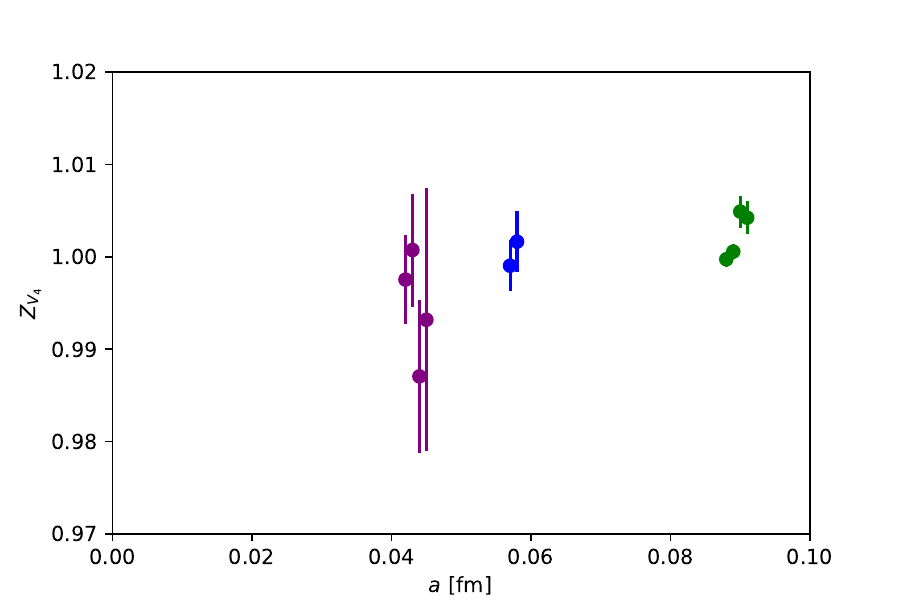}
\caption{$Z_{V_4}$ values for the heavy-light local temporal current, obtained by fitting Eq.~\eqref{eq:PCVC}, as described in the text. Results are shown for several heavy quark mass values on each ensemble. The points within each ensemble, corresponding to different heavy mass values, are offset slightly, with mass values increasing from left to right. We observe decent precision in the data, which decreases somewhat as the $H_s$ meson mass increases.
\label{fig:ZV4}}
\end{figure}

We renormalize our vector currents based on partial conservation of the vector current (PCVC). Applied to our lattice matrix elements, PCVC implies the relation
\begin{align} \label{eq:PCVC}
    Z_{V^4}(M_H - E_L) \bra{L} V^4 \ket{H}
    + Z_{V^i} \mathbf{q}\cdot \bra{L}\mathbf{V}\ket{H}
    = (m_h - m_l) \bra{L} S \ket{H} \,.
\end{align}
In the proceeding~\cite{Lytle:2023xuq}, we applied this relation in a two-step process to determine the $Z$-factors $Z_{V^4}$ and $Z_{V^i}$ for the local-temporal and one-link spatial vector currents, respectively. First zero-momentum correlators were used to determine $Z_{V^4}$, and then this was used with correlators with a chosen non-zero momentum to extract $Z_{V^i}$. Although this is straightforward, it requires a specific choice of momentum, and does not leverage information from correlators with other momenta. Here, we instead follow the strategy of Ref.~\cite{FermilabLattice:2022gku}, and fit Eq.~\eqref{eq:PCVC} using all momenta values to find best-fit values for the $Z$-factors. Sample results of this exercise are shown in Fig.~\ref{fig:ZV4} where we display $Z_{V_4}$ values for the heavy-to-light current relevant for $B_s \to K$. We observe decent precision, in most cases well below 1\%, for the $Z$-factors. We note that the precision of the determination tends to decrease for heavier physical mass values.
%----------------------------------------------------------------------
\section{Combined chiral-continuum fits}
\begin{figure}
\centering
\includegraphics[width=0.81\textwidth]{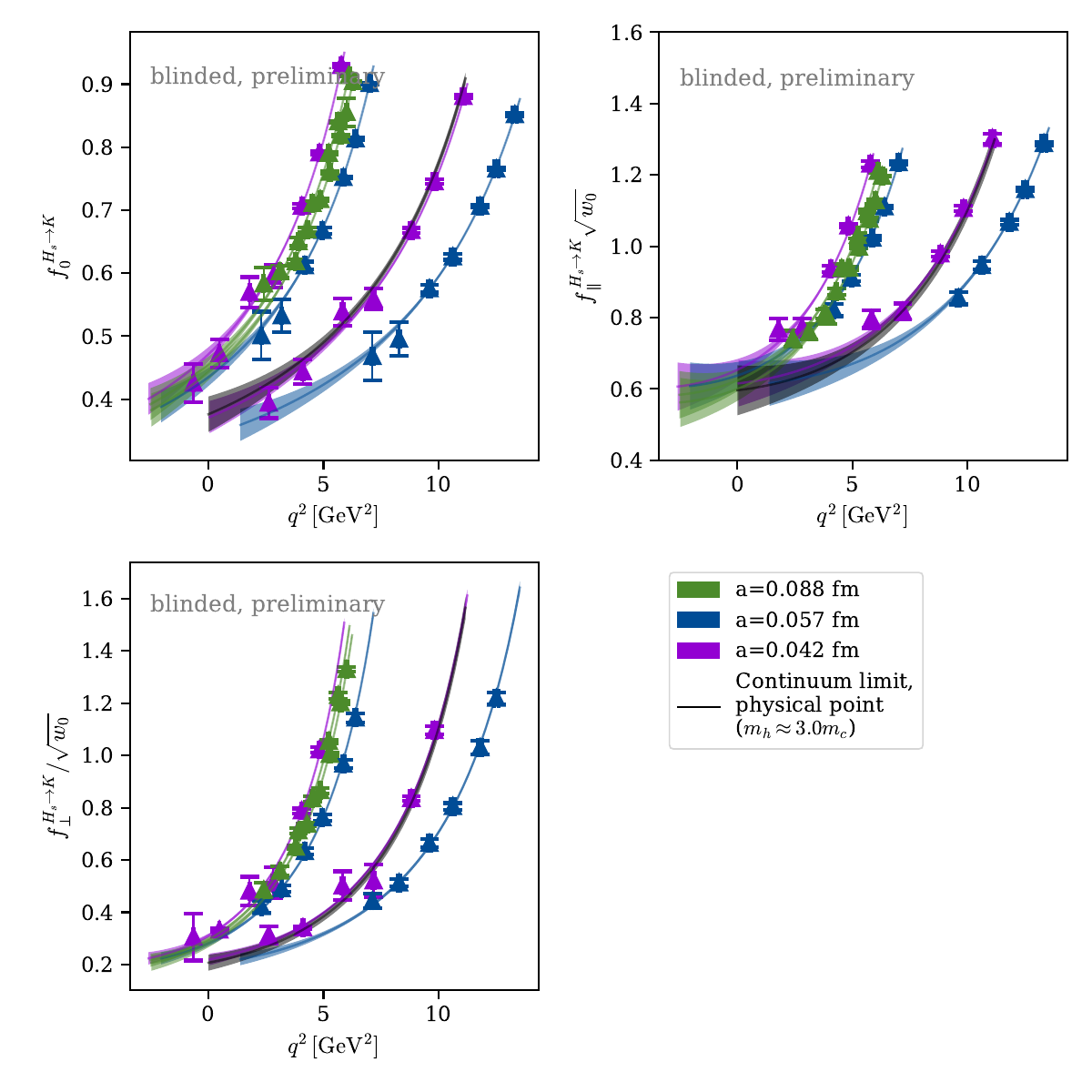}
\caption{Example, preliminary chiral fits for the $H_s \to K$ semileptonic decay, including form factors $f_0$, $f_\parallel$, and $f_\perp$. The data shown here include heavy proxy masses of $m_h \approx 2 m_c$ for 0.09 fm data (with different symbols denoting two different sets of sea masses, $m_l=0.1 m_s$, phys), $m_h \approx 2.2, 3.3 m_c$ for 0.06 fm data, and $m_h \approx 2, 3 m_c$ for 0.042 fm data. For this preliminary fit, we took the physical heavy mass to be $m_h=3 m_c$, and the continuum curve is given as a black line, nearly overlapping with $3 m_c$ data computed at our finest lattice spacing of 0.042 fm.
\label{fig:BstoK}}
\end{figure}

We can perform a chiral and continuum fit to the data, shown in Fig.~\ref{fig:BstoK}. We base our preliminary fits on hard $SU(2)$ $\chi$PT , used in the FNAL-MILC analysis of $D$-decays~\cite{FermilabLattice:2022gku}. Note that the dataset we consider here can be considered as an extension of that dataset, including quark masses approaching the physical bottom mass. In order to simultaneously fit data across a range of $m_h$ values, we will extend the chiral forms used in Ref.~\cite{FermilabLattice:2022gku} to include additional terms in an HQET-inspired expansion.

The chiral expressions from~\cite{FermilabLattice:2022gku} take the schematic form
\begin{equation}
f_{0,\parallel,\perp}(E) = \frac{c_0}{E + \Delta}(1 + \cdots + c_H \chi_{H_s} + \cdots)
\end{equation}
\begin{equation}
\Delta = \frac{M_{D^*}^2 - M_{D_s}^2 - M_K^2}{2 M_{D_s}} \,, \qquad 
\chi_{H_s} = \frac{\Lambda_{\text{HQET}}}{M_{H_s}} - \frac{\Lambda_{\text{HQET}}}{M_{D_s}^{\text{PDG}}} \,
\end{equation}
where we have broken out explicitly pieces of the expression that are modified.
We model the physical dependence of the coefficient $c_0$, similar to Ref.~\cite{Bazavov:2017lyh}, and the splitting term (treating the $D^*-D_s$ splitting to a first approximation as independent of heavy quark mass), while the mistuning term $\chi_{H_s}$ is shifted to be around the physical target final state, which we take to be at $m_h = 3 m_c$. As we refine the fits we will include the complete datasets at finer lattice spacings and shift this to $m_h = m_b$. The modifications
to the fit function are thus given by
\begin{equation}
c_0 \to c_0 + c_1 \frac{\Lambda_{\text{HQET}}} {M_{H_s}} + \cdots \,, \quad
\Delta \to \frac{M_{D^*}^2 - M_{D_s}^2 - M_K^2}{2 M_{H_s}} \text{  (1st order)} \,,
\end{equation}
\begin{equation}
\chi_{H_s} = \frac{\Lambda_{\text{HQET}}}{M_{H_s}} - \frac{\Lambda_{\text{HQET}}}{M_{H_s}^{3 m_c}} \,.
\end{equation}
The fit results evaluated at the ensemble parameters are given as colored lines matching the symbol colors in Fig.~\ref{fig:BstoK}. The continuum results, evaluated at $m_h = 3 m_c$, are shown as a black curve, and track fairly closely the 3 $m_c$ data at $a=0.042$ fm, indicating that discretization and chiral effects are fairly mild. The fits return $\chi^2/\text{dof} = 0.92,1.79,0.75$ for $f_0,f_\parallel,f_\perp$.
%----------------------------------------------------------------------
\section{Summary}
In this proceeding we provided a status update on the FNAL-MILC $B$-meson semileptonic decay calculations using HISQ quarks. We have extended our datasets to the $a=0.03$ fm ensemble with $m_l = 0.2 m_s$ light quarks in the sea. This lattice spacing can comfortably accommodate the $b$ quark, and our heaviest simulation masses straddle the $b$ mass input parameter.  Calculations on an ensemble with physical quarks and $a=0.042$ fm are underway, and preliminary results on this ensemble should be expected in the next year. Combined with results from varying sea masses on $a=0.09$ and $a=0.06$ fm ensembles, this should ensure good control over chiral effects in our final results.

We presented new, blinded results relevant for the $B_s \to K$ decay, 
which may be used to extract $\Vub$.  We presented our strategy to build on the chiral-continuum strategy used in $D$-decays, discussing preliminary chiral-continuum
fits of this data. We also presented first results for the vector-current renormalizations relevant for this decay, using a fitting strategy that leverages the complete data set. We found values for $Z_{V_4}$ that are close to 1, and generally have sub-percent precision, although the precision decreases as heavier masses are considered.

%----------------------------------------------------------------------
\section{Acknowledgments}
\vskip -0.25cm
This material is based upon work supported
by the U.S. Department of Energy, Office of Science under grant Contract Numbers DE-SC0015655 (A.L., A.X.K.), DE-SC0010120 (S.G.), DE-SC0011090 (W.J.), DE-SC0021006 (W.J.), and DE-SC0009998 (J.L.); 
by the Simons Foundation under their Simons Fellows in Theoretical Physics program (A.X.K.);
by the U.S. National Science Foundation under Grants No.\ PHY17-19626 and PHY20-13064 (C.D., A.V.);
by MCIN/AEI/10.13039/501100011033/FEDER, UE under Grants No.\ PID2019-106087GB-C21 and PID2022-140440NB-C21 (E.G.);
by the Junta de Andalucía (Spain) under Grant No.\ FQM-101 (E.G.);
and by AEI (Spain) under Grant No.\ RYC2020-030244-I / AEI / 10.13039/501100011033 (A.V.).
This document was prepared using the resources of the Fermi National Accelerator Laboratory (Fermilab), a U.S. Department of Energy, Office of Science, HEP User Facility.
Fermilab is managed by Fermi Research Alliance, LLC (FRA), acting under Contract No. DE-AC02-07CH11359.

Computations for this work were carried out in part on facilities of the USQCD Collaboration, which are funded by the Office of Science of the U.S. Department of Energy.
%INCITE/ALCF/OLCF
An award of computer time was provided by the Innovative and Novel Computational Impact on Theory and Experiment (INCITE) program. This research used resources of the Argonne Leadership Computing Facility, which is a DOE Office of Science User Facility supported under contract DE-AC02-06CH11357. This research also used resources of the Oak Ridge Leadership Computing Facility, which is a DOE Office of Science User Facility supported under Contract DE-AC05-00OR22725.
%NERSC
This research used resources of the National Energy Research Scientific Computing Center (NERSC), a U.S. Department of Energy Office of Science User Facility located at Lawrence Berkeley National Laboratory, operated under Contract No. DE-AC02-05CH11231.
%ALCC???
The authors acknowledge support from the ASCR Leadership Computing Challenge (ALCC) in the form of time on the computers Summit and Theta.
%TACC
The authors acknowledge the \href{http://www.tacc.utexas.edu}{Texas Advanced Computing Center (TACC)} at The University of Texas at Austin for providing HPC resources that have contributed to the research results reported within this paper.
This research is part of the Frontera computing project at the Texas Advanced Computing Center. Frontera is made possible by National Science Foundation award OAC-1818253~\cite{Frontera}.
%XSEDE
This work used the Extreme Science and Engineering Discovery Environment (XSEDE), which is supported by National Science Foundation grant number ACI-1548562.
This work used XSEDE Ranch through the allocation TG-MCA93S002~\cite{XSEDE}.

%----------------------------------------------------------------------

\end{document}